\documentclass[aps,prl,showpacs,twocolumn]{revtex4-1}

\usepackage{amsmath} 
\usepackage{graphicx}

\newcommand{\rms}{\rm\scriptscriptstyle} 

\begin{document}

\title{Classical Driven Transport in Open Systems with Particle
  Interactions\\ and General Couplings to Reservoirs}

\author{Marcel Dierl} \author{Philipp Maass} \author{Mario Einax}

\affiliation{Fachbereich Physik, Universit\"at Osnabr\"uck,
  Barbarastra\ss e 7, 49076 Osnabr\"uck, Germany}

\date{\today}

\begin{abstract}

  We study nonequilibrium steady states of lattice gases with
  nearest-neighbor interactions that are driven between two
  reservoirs. Density profiles in these systems exhibit oscillations
  close to the reservoirs. We demonstrate that an approach based on
  time-dependent density functional theory copes with these
  oscillations and predicts phase diagrams of bulk densities to a good
  approximation under arbitrary boundary-reservoir couplings. The
  minimum or maximum current principles can be applied only for
  specific bulk-adapted couplings. We show that they generally fail to
  give the correct topology of phase diagrams but can still be useful
  for getting insight into the mutual arrangement of different phases.

\end{abstract}


\maketitle

Lattice gases provide useful models for investigating driven particle
transport in biological, chemical and physical systems
\cite{Hille:2001,Nitzan:2006}. They are used to study Brownian
ratchets and motors \cite{Hanggi:2009, Einax/et_al_2:2010}, organic
photovoltaic cells \cite{Einax/et_al_3:2011}, and traffic on networks
\cite{Neri:2011}, to name only some recent applications. In these
examples, the asymmetric simple exclusion process (ASEP) appears as a
basic building block for the description of low-dimensional
transport. As such it has developed into one of the standard models
for investigating nonequilibrium steady states (NESS). In the ASEP,
particles hop under the influence of a bias field and cannot occupy
the same place. For open boundary conditions, these biased systems
show intriguing phenomena, like boundary-induced phase transitions
\cite{Krug:1991}, emergence of shock fronts \cite{Kolomeisky:1998},
and self-organized pattern formation \cite{Boal:1991}. To explain
these phenomena, it is generally sufficient to consider the totally
asymmetric simple exclusion processes (TASEPs), where the particle
transport in the bias direction is unidirectional.

Much progress has been made in the past to understand ASEPs and TASEPs
(for reviews, see Refs.~\cite{Derrida:1998,Golinelli:2006,
  Blythe:2007}), but only a few studies so far have addressed TASEPs
with particle-particle interaction going beyond hard-core repulsions
\cite{Katz:1984,Krug:1991,Popkov:1999,Antal:2000,Hager:2001,Dierl:2011}. While
in this case a complete description of the NESS seems to be out of
reach for open systems, it has been shown that the minimum/maximum
current principles \cite{Krug:1991,Popkov:1999}, or the domain-wall
(shock-front) theory \cite{Kolomeisky:1998,Popkov:1999,Bauer:1999},
can be used in a specific model setup to predict phase diagrams of the
bulk density. This setup requires that the relations between
correlation functions and densities close to the system-reservoir
boundaries are the same as in the bulk. To ensure this, a particular
way of particle injection and ejection has to be taken, which was used
in Refs.~\cite{Popkov:1999,Antal:2000,Hager:2001,Dierl:2011}.

However, realistic reservoirs are not of that type but are specified
by only a few control parameters, as, for example, temperature and
chemical potential.  It is therefore needed to develop methods that
can deal with such situations both on basic reasons and in view of
various applications. For example, in driven transport through
molecular bridges (e.g., in the incoherent limit for weak coupling to
leads) \cite{Nitzan:2001,Harbola:2006}, one does not specify a
complicated bath-system coupling, but is led by the fact that the
typical relaxation dynamics in the bath is much faster than in the
system.  Accordingly, the baths are supposed to be in equilibrium and
characterized essentially by their chemical potential.

\begin{figure}[t!]
\centering \includegraphics[width=0.38\textwidth]{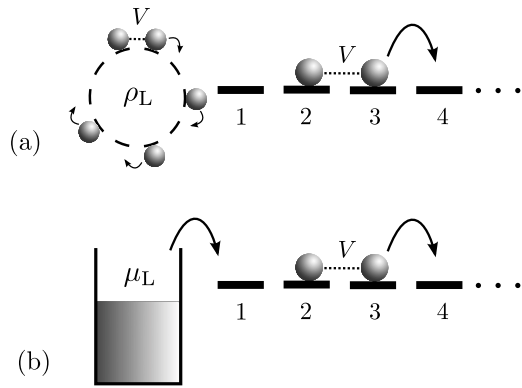}
\caption{Sketch of the model and illustration of the
  particle injection for (a) the bulk-adapted, and (b)
  the equilibrated-bath couplings. Right reservoirs are not
  shown.}
 \label{fig:fig1}
\end{figure}

In this Letter we consider a TASEP with nearest-neighbor interactions
as sketched in Fig.~\ref{fig:fig1} and show that the couplings of the
system to the reservoirs have a strong influence on the density
profiles in the nonequilibrium steady state (NESS). In general,
oscillations of these profiles occur close to the boundaries, similar
as they are known for equilibrium systems.  We present a theory that
is able to cope with these oscillations and to describe the density
profiles to a good approximation. Based on the theory, the phase
diagram of the bulk density can be determined. Our theoretical
approach provides a general method to predict such phase diagrams for
driven systems with interactions under general boundary-reservoir
couplings.

To demonstrate our theoretical approach we consider as an example the
following specific model, see Fig.~\ref{fig:fig1}: Particles with
repulsive nearest-neighbor interaction of strength $V>0$ perform a
unidirectional hopping motion between neighboring sites of a lattice
with $N$ sites and are thereby transported from a left to a right
particle reservoir. The microstate of the system is specified by the
set of occupation numbers $\boldsymbol n = \{ n_i \}$, $i=1,\ldots,N$,
where $n_i = 0$ or $1$ if the corresponding site is vacant or occupied
by a particle.  The jump rate of a particle from a site $i$ to a
vacant neighboring site $(i+1)$ is $\Gamma_i=\exp(-\Delta E/2)$, where
$\Delta E$ is the energy difference (in units of the thermal energy)
between the final and initial state after and before the jump.

When $V>V_{\rm c}=2.89$, the bulk current-density relation of this
model exhibits a double-hump structure with two maxima at densities
$\rho_{{\rm max},1}$ and $\rho_{{\rm max},2}$, and a minimum in
between at $\rho=0.5$ \cite{Dierl:2011}. The bulk dynamics are
considered to apply to jumps from sites $i=2,\ldots,(N-2)$, while
couplings to the reservoirs are taken into account for injections from
and ejections to the reservoirs. In addition we need to specify rates
for jumps from sites $i=1$ and $(N-1)$, where, if one adopted the bulk
rates, a nearest- and a next-nearest-neighbor site would be missing to
the left and right, respectively.  As sketched in Fig.~\ref{fig:fig1},
two cases of boundary couplings are considered. These couplings are
referred to as ``bulk-adapted'' and ``equilibrated-bath'' coupling and
explained for the left reservoir in the following. Corresponding
couplings are applied to the right reservoir.

For the equilibrated-bath coupling [Fig.~\ref{fig:fig1}(b)], the left
reservoir is considered to be an equilibrated ideal Fermi gas with a
chemical potential $\mu_{\rms L}$, corresponding to a reservoir
density $\rho_{\rms L}=1/[\exp(-\mu_{\rms L})+1]$. Accordingly, we
write for the rate $\Gamma_{\rms L}$ of particle injection
$\Gamma_{\rms L}(n_2)=\rho_{\rms L}\exp[(\mu_{\rms L}-n_2V)/2]$, and
$\Gamma_1(n_3)=\exp(-n_3V/2)$ for the jump rate from site one.

The bulk-adapted coupling [Fig.~\ref{fig:fig1}(a)] is arranged in such
a way that the system can be viewed as being continued into a
reservoir with density $\rho_{\rms L}$, corresponding to relations
between correlation functions \ $\langle n_in_j\ldots\rangle$, and
densities $\rho_i=\langle n_i\rangle$ as in the bulk, where
$\langle\ldots\rangle$ denotes an average over the distribution of
microstates in the NESS. In a closed bulk (ring) system, when an
initial configuration $\{n_{i+1}=0,n_{i+2}\}$ would be given, two
rates are possible for a particle jump from site $i$ (i.e.\ $n_i=1$):
$\Gamma_i=\exp(-n_{i+2}V/2)$, if $n_{i-1}=0$, while
$\Gamma_i=\exp[(1-n_{i+2})V/2]$, if $n_{i-1}=1$. For given
$\{n_{i+1}=0,n_{i+2}\}$, let us denote by $p(01|0n_{i+2})$ and
$p(11|0n_{i+2})$ the conditional probabilities for the configurations
$\{n_{i-1},n_i\}=\{0,1\}$ and $\{n_{i-1},n_i\}=\{1,1\}$ to occur in
the NESS of a closed bulk system with density $\rho_{\rms L}$ and
interaction $V$, respectively. The injection rate $\Gamma_{\rms
  L}(n_2)$ then results from a weighting of rates with the
probabilities $p(01|0n_2)$ and $p(11|0n_2)$ corresponding to virtual
configurations $\{n_{-1}=0,n_{0}=1,n_{1}=0,n_{2}\}$ and
$\{n_{-1}=1,n_{0}=1,n_{1}=0,n_{2}\}$ at the boundaries, i.e.,
$\Gamma_{\rms
  L}(n_2)=p(01|0n_2)\exp(-n_2V/2)+p(11|0n_2)\exp[(1-n_2)V/2]$.
Analogously, $\Gamma_1(n_3)=p(0|10n_3)\exp(-n_3V/2)+
p(1|10n_3)\exp[(1-n_3)V/2]$. For a practical implementation of a
corresponding kinetic Monte Carlo (KMC) simulation, the conditional
probabilities $p(.|.)$ are determined from separate KMC simulations of
periodic ring systems with a particle density $\rho=\rho_{\rms L}$ as
indicated in Fig.~\ref{fig:fig1}(a). The rates for the two models are
summarized in \cite{suppl:rates}.

\begin{figure}[t!]
\includegraphics[width=0.45\textwidth]{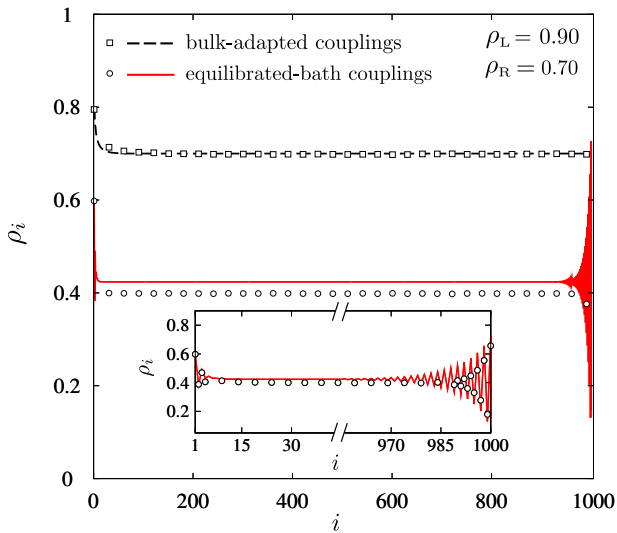}
\caption{(Color online) Steady-state density profiles for $V=2V_{\rm
    c}$ from KMC simulations (symbols) in comparison with the TDFT
  predictions (lines).  For the KMC results averages were performed
  over $10^9$ particle jumps in the steady state. The inset shows the
  density profile for the equilibrated-bath couplings close to the
  left and right boundary, respectively.}
\label{fig:fig2}
\end{figure}

KMC simulations have been carried out for both the bulk-adapted and
the equilibrated-bath coupling and results are shown by the squares
and circles in Fig.~\ref{fig:fig2} for $V=2V_{\rm c}$, $\rho_{\rms
  L}=0.9$ and $\rho_{\rms R}=0.7$. As a consequence of the specific
arrangements in the bulk-adapted case, monotonously varying density
profiles are obtained, similar as in the TASEP with hard-core
repulsion only.  The value of the bulk density $\rho_{\rms
  B}\cong0.70$ agrees with that predicted from applying the maximum
current principle to the bulk current-density relation. For the
equilibrated-bath coupling by contrast, pronounced oscillations appear
at the boundaries. This implies that methods relying on the bulk
current-density relation (minimum/maximum current principles,
shock-front/domain-wall theories) cannot be applied any more to
predict the bulk densities. In fact, Fig.~\ref{fig:fig2} shows that a
value $\rho_{\rms B}\cong0.40$ is obtained for the equilibrated-bath
coupling, which differs from that for the bulk-adapted coupling.

The question is whether a theory can successfully account for the bulk
densities and associated phase diagrams for interacting driven
particle systems that generally will show density oscillations at the
boundaries. Mean-field theories with simple factorization schemes, as,
for example, $\langle n_in_j\rangle\simeq\langle n_i\rangle\langle
n_j\rangle$ fail for the model introduced here, since they are not
even capable to predict the double-hump structure in the bulk
current-density relation.  We now show that the application of the
time-dependent density functional theory (TDFT) presented in
\cite{Heinrichs:2004,Dierl:2011} can deal with the complications
associated with the density oscillations.  In this approach the
average current $j_i$ from site $i$ to site $(i+1)$ in the bulk
($i=2,\ldots,(N-2)$) is given by
\begin{align}
\label{eq:current-tdft}
j_i&=\left[(\rho_{i+2} - C_{i+1})e^{-V/2} + \tilde \rho_{i+1} -
  \rho_{i+2} + C_{i+1}\right]\nonumber\\
&\hspace{1.3em}{}\times
\frac{\rho_i - C_i}{\rho_i \tilde \rho_{i+1}}\left[\rho_i - C_{i-1} +
  e^{V/2}C_{i-1}\right]\, ,
\end{align}
where $\rho_i = \langle n_i \rangle$ and $\tilde \rho_i = 1- \langle
n_i \rangle = 1-\rho_i$ are the particle and hole density,
respectively, and $C_i=\langle n_i n_{i+1}\rangle$ is the two-point
correlation function \cite{Buschle:2000},
\begin{equation}
  C_i = \exp\left(-V\right) \frac{\left(\rho_i- C_i\right)
  \left(\rho_{i+1}- C_i\right) } {1 - \rho_i - \rho_{i+1} + C_i}\,.
  \label{eq:correlator}
\end{equation}
This expression can be explicitly solved to yield functions
$C_i=C_i(\rho_i,\rho_{i+1})$ and in this way, the bulk currents $j_i$
in Eq.~\eqref{eq:current-tdft} become functionals of the density
profile $\boldsymbol{\rho}(t) =\{\rho_i(t)\}$.

For the bulk-adapted couplings, Eq.~\eqref{eq:current-tdft} applies
also at the boundaries, that means for $i=0$, 1, $(N-1)$ and $N$
\cite{comm:j}.  For the equilibrated-bath couplings, however,
different functional structures are obtained:
\begin{subequations}
\label{eq:bc}
\begin{align}
 \label{eq:bc1}
 j_0 &= \rho_{\rm L}e^{\mu_{\rm
     L}/2}\left[\left(\rho_2-C_1\right)e^{-V/2} + \tilde \rho_1 -
   \rho_2 + C_1\right]\,,\\
 \label{eq:bc2}
 j_1 &= \frac{\rho_1 - C_1}{\tilde
   \rho_2}\left[\left(\rho_3-C_2\right)e^{-V/2} + \tilde \rho_2 -
   \rho_3 + C_2\right]\,,\\
 \label{eq:bc3}
 j_{N-1} &= \frac{\rho_{N-1}-C_{N-1}}{\rho_{N-1}} \left[C_{N-2}e^{V/2}
   + \rho_{N-1} - C_{N-2}\right]\,,\\
  \label{eq:bc4}
 j_N &= \tilde \rho_{\rm R}e^{-\mu_{\rm R}/2}\left[C_{N-1}e^{V/2} +
   \rho_N - C_{N-1}\right]\, .
\end{align}
\end{subequations}

To compare the predictions of this theory with the KMC results we have
integrated numerically the coupled set of rate equations ${\rm
  d}\rho_i(t)/{\rm d}t=j_{i-1}(t) - j_i(t)$ ($i=1,\,\ldots, N$) with
the currents given by \eqref{eq:current-tdft}-\eqref{eq:bc} and
evaluated the long-time limit to analyze the NESS. As shown in
Fig.~\ref{fig:fig2}, the density profile from the TDFT for the
bulk-adapted coupling agrees well with KMC results. Surprisingly, also
the oscillations of the KMC density profiles for the equilibrated-bath
couplings are closely reproduced by the TDFT, see the inset of
Fig.~\ref{fig:fig2}. Moreover, the TDFT gives a value $\rho_{\rms
  B}\cong0.42$ that is only slightly larger than the corresponding KMC
value $\rho_{\rms B}\cong0.40$. For the steady-state currents we find
$j_{\rms NESS}=0.174$ in the KMC and $j_{\rms NESS}=0.149$ in the
TDFT.

\begin{figure}[t!]
\centering
\includegraphics[width=0.43\textwidth]{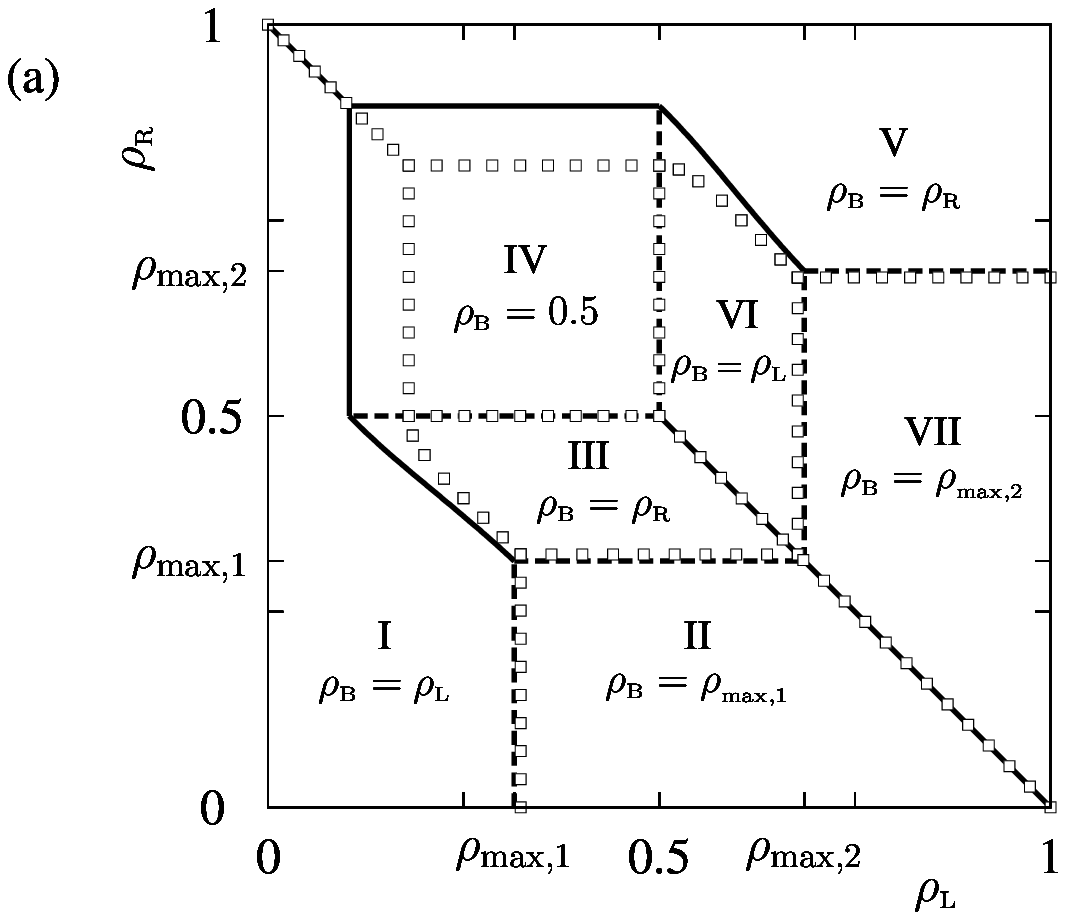}\vspace*{7ex}
\centering \includegraphics[width=0.43\textwidth]{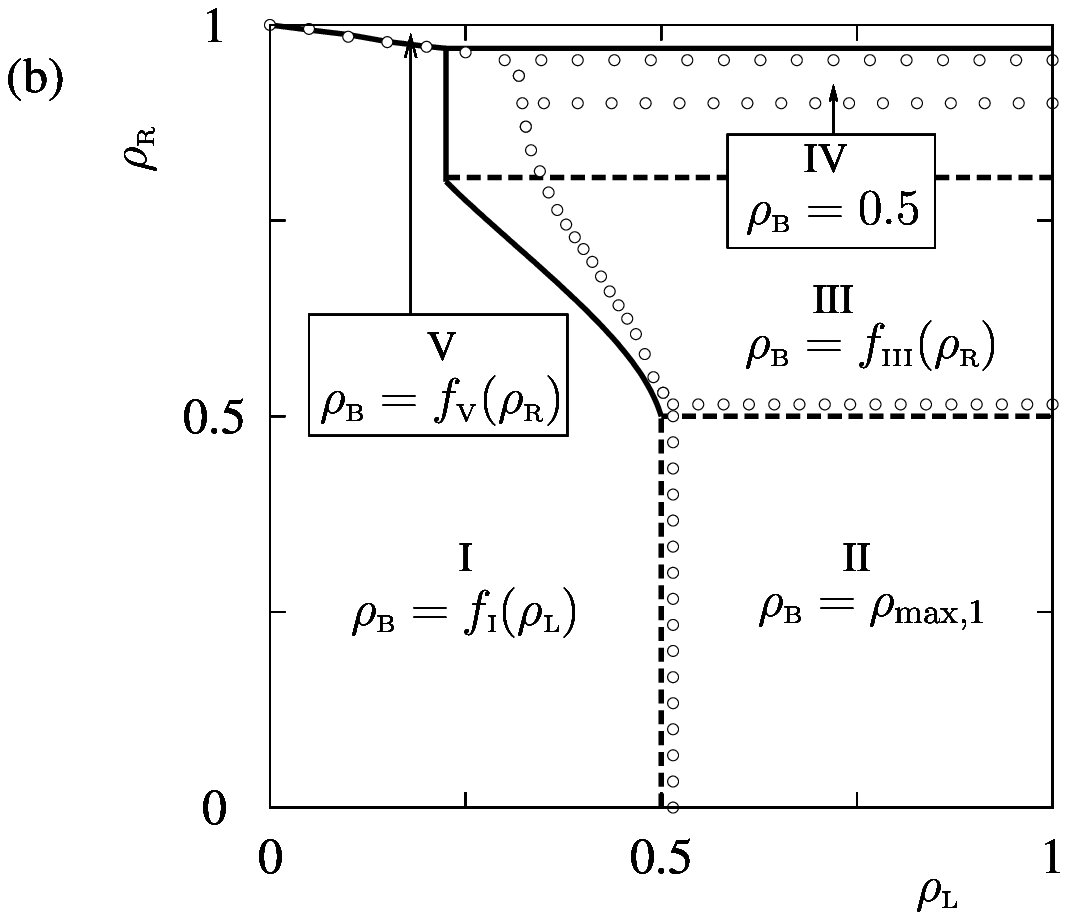}
\caption{Phase diagrams of the NESS for (a) the bulk-adapted and (b)
  the equilibrated-bath couplings at $V =2V_{\rm c}$.  KMC results for
  the phase transitions are marked by the symbols and TDFT results by
  the lines (solid lines for first-order and dashed lines for
  second-order phase transitions).  In (a) seven phases are obtained, where
  the bulk density $\rho_{\rms B}$ equals either the reservoir
  densities $\rho_{\rms L}$ or $\rho_{\rms R}$, or the value $0.5$ in
  the minimum current phase, or the two possible values $\rho_{{\rm
      max},1}$ or $\rho_{{\rm max},2}$ in the maximum current phases.
  In (b) a maximum current phase with $\rho_{\rms B}=\rho_{{\rm
      max},1}$, a minimum current phase with $\rho_{\rms B}=0.5$, and
  three phases with reservoir-controlled densities $\rho_{\rms
    B}=f_{\rms I}(\rho_{\rms L})$, $\rho_{\rms B}=f_{\rms
    III}(\rho_{\rms R})$, and $\rho_{\rms B}=f_{\rms V}(\rho_{\rms
    R})$ are obtained, where the different functions of $\rho_{\rms
    L}$ or $\rho_{\rms R}$ are determined from the simulated or
  calculated density profiles. The labeling by roman numbers in (a)
  and (b) has been chosen in such a manner that corresponding phases
  have equal numbers.}
\label{fig:fig3}
\end{figure}

With the density profiles determined from either KMC or TDFT we can
identify the singularities in the dependence of the bulk density on
the reservoir densities $\rho_{\rms L}$ and $\rho_{\rms R}$.
Corresponding transition lines are shown in the phase diagram of
Fig.~\ref{fig:fig3} for $V=2V_{\rm c}$, and for (a) the bulk-adapted
and (b) the equilibrated-bath couplings. Overall, the TDFT accounts
well for the phase transitions as determined from the KMC
simulations. Due to the construction of the bulk-adapted couplings,
the knowledge of the exact bulk current-density relation in the NESS
would allow one to determine exactly the phase diagram by applying the
minimum/maximum current principles. In this way seven phases are
identified in Fig.~\ref{fig:fig3}(a). Since the TDFT does not yield
the bulk current-density relation in the NESS exactly, small
deviations are seen in Fig.~\ref{fig:fig3}(a) between KMC and TDFT
results.

The strongly different diagram in Fig.~\ref{fig:fig3}(b) shows that
the minimum/maximum current principles are no longer successful. By
contrast, the TDFT accounts well for the five phases identified in the
KMC simulations. There are three phases, where $\rho_{\rms B}$ is some
function of either $\rho_{\rms L}$ or $\rho_{\rms R}$, and a maximum
current phase with $\rho_{\rms B}=\rho_{{\rm max},1}$ and a minimum
current phase with $\rho_{\rms B}=0.5$. Generally, $\rho_{\rms B}$
must be either determined by $\rho_{ \rms L}$ or $\rho_{\rms R}$, or
by the extrema in the bulk current-density relation, since the
minimum/maximum current principles still apply in the inner bulk
regions, where the density profiles are monotonously varying. The
absence of a maximum current phase with $\rho_{\rms B}=\rho_{{\rm
    max},2}$ for the equilibrated-bath couplings means that also the
topology of the phase diagram in Fig.~\ref{fig:fig3}(b) is changed
compared to that in Fig.~\ref{fig:fig3}(a). It can be shown that this
change of topology is generally possible and that there are no further
hidden phases in addition to the ones shown in Fig.~\ref{fig:fig3}(b)
\cite{Dierl/etal:2011}. Those phases, which appear in
Fig.~\ref{fig:fig3}(b), have a connection to the ones in
Fig.~\ref{fig:fig3}(a) in the sense that their mutual arrangement
remains the same.

In conclusion we have shown that the theoretical approach based on the
TDFT can cope with the problem of driven lattice gases with extended
interactions, where oscillations in densities and correlation
functions naturally occur at the boundaries. The oscillations imply
that the minimum/maximum current principles are no longer sufficient,
since they can only be applied in an inner bulk region, where the
density profile varies monotonously. When, for unmodified bulk
dynamics, deliberately changing the boundary couplings to the
bulk-adapted ones in order to enable the use of the minimum/maximum
current principles, the currents and bulk densities in the NESS as
well as the associated phase diagrams are strongly influenced. The
principles are nevertheless useful to justify the identification of
possible phases on the basis of a known bulk current-density
relation. The investigation of the bulk-adapted couplings can be
helpful to predict the mutual arrangement of those phases that appear
for the boundary couplings of interest.

We believe that our theory can be useful to explore a wider range of
driven open systems which are of importance for many biological
processes and electronic transport phenomena in small molecular
devices that can be treated within the incoherent classical limit.

\begin{acknowledgments}
We thank W.~Dieterich and A.~Nitzan for very illuminating discussions
concerning this work.
\end{acknowledgments}


\bibliography{lit}

\begin{thebibliography}{25}
\bibitem [{com()}]{comm:j}%
  \BibitemOpen
  \href@noop {} {}\bibinfo {note} {Note that $j_0$ and $j_N$ refer to injection
  and ejection currents connected with the reservoirs. Local densities $\rho_i$
  with indices $i\le0$ and $i\ge(N+1)$ appearing on the right hand side of
  Eq.~\eqref{eq:current-tdft} are equal to the reservoir densities $\rho_{\rms
  L}$ and $\rho_{\rms R}$, respectively.}\BibitemShut {Stop}%
\end{thebibliography}
\bibliographystyle{apsrev4-1}

\end{document}